\documentclass[conference]{IEEEtran}
\usepackage{graphicx}
\usepackage{amsbsy}
\usepackage{bm}
\usepackage{lettrine}
\usepackage{amsfonts}
\usepackage{csquotes}
\usepackage{epstopdf}
\usepackage{cite}
\usepackage[lined,ruled,commentsnumbered]{algorithm2e}
\usepackage{amsmath,amssymb}
\usepackage{amsthm}
\DeclareMathOperator*{\argmin}{argmin}

\begin{document}

\title{Iterative Matrix Inversion Based Low Complexity Detection in Large/Massive MIMO Systems }

\author{
\IEEEauthorblockN{Vipul Gupta$^*$, Abhay Kumar Sah$^\dag$ and A. K. Chaturvedi$^\ddag$}
\IEEEauthorblockA
{Department of Electrical Engineering,\\
Indian Institute of Technology Kanpur\\
Kanpur, India 208016\\
Email: \{vipgupta$^*$, abhaysah$^\dag$, akc$^\ddag$\}@iitk.ac.in}
}

\maketitle

\begin{abstract} 
Linear detectors such as zero forcing (ZF) or minimum mean square error (MMSE) are imperative for large/massive MIMO systems for both the downlink and uplink scenarios. However these linear detectors require matrix inversion which is computationally expensive for such huge systems. In this paper, we assert that calculating an exact inverse is not necessary to find the ZF/MMSE solution and an approximate inverse would yield a similar performance. This is possible if the quantized solution calculated using the approximate inverse is same as the one calculated using the exact inverse. We quantify the amount of approximation that can be tolerated for this to happen. Motivated by this, we propose to use the existing iterative methods for obtaining low complexity approximate inverses. We show that, after a sufficient number of iterations, the inverse using iterative methods can provide a similar error performance. In addition, we also show that the advantage of using an approximate inverse is not limited to linear detectors but can be extended to non linear detectors such as sphere decoders (SD). An approximate inverse can be used for any SD that requires matrix inversion. We prove that application of approximate inverse leads to a smaller radius, which in turn reduces the search space leading to reduction in complexity.  Numerical results corroborate our claim that using approximate matrix inversion reduces decoding complexity in large/massive MIMO systems with no loss in error performance.

\end{abstract}

\section{Introduction}

With growing demand for high throughput, Mutiple-Input-Multiple-Output (MIMO) systems with large/massive number of antennas are expected to become an indispensable part of fifth generation wireless technology \cite{Larsson14, Heath14}. It employs a large number of antennas at the base station (of the order of hundreds) that operate to serve relatively fewer users. However, we know that as the number of antennas grow, the complexity of detection algorithms increases \cite{Lu14}. Thus, there is need for techniques which, while exploiting the extra degrees of freedom, are able to decode the transmitted signal efficiently in terms of error performance and complexity.

In the literature, Zero Forcing (ZF) and Minimum Mean Square Error (MMSE) have commonly been used as precoders in a massive MIMO downlink \cite{Hoydis13,Haoming14} and as decoders in a massive MIMO uplink. Even the complex decoders for uplink transmission also require the computation of ZF/MMSE solution. For example, neighborhood search based algorithms \cite{Self_GC15, Datta11} or sparsity based detectors \cite{Liu15, Choi14} use such linear detectors for initialization. Calculating a ZF or an MMSE solution requires inversion of a matrix. However, finding an inverse is computationally expensive, especially when large number of antennas are employed.

In this paper, we argue that an approximate matrix inverse suffices for finding a ZF/MMSE solution. In other words, usage of an approximate inverse does not compromise the quality of a ZF/MMSE solution. Since the solution obtained using linear detectors anyway needs to be quantized, it is clear that there is a scope for using an approximate inverse as long as the quantized solution remains unchanged. We derive bounds on the approximation such that the ZF/MMSE solutions from the exact and approximate inverses are same. 
Further, we show that the advantages of using an approximate inversion are not limited to linear detectors. 
Thus, a class of Sphere Decoding (SD)  algorithms \cite{ref4} require the ZF solution for computing the Babai Radius (BR) \cite{Hassibi05, babai2}, consequently requiring matrix inversion. 
Hence, one can think of utilizing an approximate matrix inverse even in complex decoding schemes like SD.

In this work, we propose the application of an approximate inverse to compute the BR for usage in SD. The approximate inverse has two advantages. Firstly, it reduces the complexity of matrix inversion. But secondly, and more importantly, we prove that it results in a smaller BR. This is a bigger advantage as complexity of decoding in such SD algorithms is largely governed by the choice of BR. Simulations results for large/massive MIMO systems corroborate that the proposed SD provides a low complexity solution with no loss in error performance.  

 
\section{System Model}

Consider a massive MIMO downlink with $N$ transmit antennas at the base station and $K$ users, each with a single receive antenna. Such a system can be represented by
\begin{eqnarray}\label{sys_model_downlink}
\mathbf{y}_d = \mathbf{H}_d\mathbf{s}_d + \mathbf{n}_d,
\end{eqnarray}
where $\mathbf{s}_d = \mathbf{Wx}_d$, $\mathbf{W}$ is the linear precoder such as ZF or MMSE and $\mathbf{x}_d$ is the $N$ dimensional signal vector transmitted from the base station. Each element in $\mathbf{x}_d$ is drawn from a set $\Omega$, all entries of which belong to an $M$-QAM constellation, with average symbol energy $E_s$. $\mathbf{H}_d$ represents the $K\times N$ i.i.d. channel matrix with zero mean and unit variance and $\mathbf{n}_d$ is an i.i.d. zero mean Gaussian noise vector with dimension $K\times 1$ and variance $N_{0}$. The $i$-th entry of the vector $\mathbf{y}_d$, $y_{i,d}$, is the signal intended for the $i$-th user, for $i=1,2,\cdots, K$. 


Similarly, in the case of uplink, the system can be represented by 
\begin{eqnarray}\label{sys_model_uplink}
\mathbf{y}_u = \mathbf{H}_u\mathbf{x}_u + \mathbf{n}_u,
\end{eqnarray}
where $\mathbf{x}_u$ is the $K$ dimensional transmitted signal vector whose $i$-th entry is the symbol transmitted by the $i$-th user, for $i=1,2,\cdots, K$. Again, each element in $\mathbf{x}_u$ is drawn from the set $\Omega$, with average symbol energy $E_s$. Similarly, $\mathbf{H}_u$ is the $N\times K$ i.i.d. channel matrix with each coefficient having zero mean and unit variance. The noise vector $\mathbf{n}_u$ is i.i.d. $N\times 1$ Gaussian with each element having zero mean and variance $N_{0}$, and $\mathbf{y}_u$ is the $N$ dimensional received signal vector at the base station. 
This results in $KE_s/N_0$ Signal-to-Noise Ratio (SNR) at each receive antenna.

\section{A Linear Detector using Approximate Matrix Inverse}

Linear detectors such as ZF and MMSE are useful for both the uplink and downlink (as a precoder) in massive MIMO systems. The expressions for these detectors can be expressed as
\begin{eqnarray}
\mathbf{x}_{\text{ZF}} &=& \left\lceil (\mathbf{H}^H\mathbf{H})^{-1}\mathbf{H}^H\mathbf{y}\right\rfloor\\
\mathbf{x}_{\text{MMSE}} &=& \left\lceil (\mathbf{H}^H\mathbf{H} + \frac{N_0}{E_s}\mathbf{I}_{K})^{-1}\mathbf{H}^H\mathbf{y}\right\rfloor,
\end{eqnarray}
where $\lceil\cdot\rfloor$ quantization operator to the set $\Omega$ and $\mathbf{H}$ is the $N \times K$ channel matrix. Quantization allows us to use an approximate inverse instead of exact inverse while giving the same ZF/MMSE solution. Since the operations are similar in both the uplink and downlink scenarios, we consider only the uplink scenario for the analysis. For  notational simplicity, we have removed the subscripts here onwards. 

Let us define the error in the approximation of the inverse of a matrix $\mathbf{C}$ as $\mathbf{E} = \mathbf{\widetilde{C}} - \mathbf{C}^{-1}$, where $\mathbf{C} = \mathbf{H}^H\mathbf{H}$ is a $K \times K$ matrix that needs to be inverted and $\mathbf{\widetilde C}$ is its approximate inverse.  Also, define $\mathbf{g} = \mathbf{H}^H\mathbf{y}.$

\subsection{A Bound on the Acceptable Error in the Matrix Inverse}
We will consider an approximate matrix inverse good if the ZF solution calculated through it is equal to that calculated through the exact inverse. In this section, we evaluate a bound on the error which can be tolerated in the computation of an approximate matrix inverse. 

For the ZF solutions calculated using the exact and approximate matrix inverses to be equal, the following equality must be satisfied 
\begin{eqnarray}\label{condition}
\underset{\mathbf{x} \in \Omega}{\argmin}\|\mathbf{x}-\mathbf{C}^{-1}\mathbf{H}^H\mathbf{y}\|^2 &=& \underset{\mathbf{x}\in \Omega}{\argmin}\|\mathbf{x}-\mathbf{\widetilde C}\mathbf{H}^H\mathbf{y}\|^2 \nonumber \\
\Rightarrow \underset{\mathbf{x} \in \Omega}{\argmin}\|\mathbf{x}-\mathbf{C}^{-1}\mathbf{g}\|^2 &=& \underset{\mathbf{x}\in \Omega}{\argmin}\|\mathbf{x}-\mathbf{\widetilde C}\mathbf{g}\|^2 \nonumber \\
\Rightarrow \underset{\mathbf{x} \in \Omega}{\argmin}\|\mathbf{x}-\mathbf{C}^{-1}\mathbf{g}\|^2 &=& \underset{\mathbf{x}\in\Omega}{\argmin}\|\mathbf{x}-\mathbf{C}^{-1}\mathbf{g} - \mathbf{E}\mathbf{g}\|^2.\nonumber\\
\end{eqnarray}

Let the solution of the L.H.S. of \eqref{condition} be $\mathbf{x}_{\text{ZF}}$ and let $\mathbf{z} = \mathbf{x}_{\text{ZF}} - \mathbf{C}^{-1}\mathbf{g}.$
Therefore, (\ref{condition}) will be satisfied if the following inequalities are satisfied by the error matrix $\mathbf{E}$ (a sufficient condition)
\begin{eqnarray}
-\frac{d_{\min}}{2} < \Re(z_i - \sum_{j=1}^K E_{ij}g_j) < \frac{d_{\min}}{2} \label{eq1} \\
-\frac{d_{\min}}{2} < \Im(z_i - \sum_{j=1}^K E_{ij}g_j) < \frac{d_{\min}}{2} \label{eq2},
\end{eqnarray}
$\forall~ i= 1, 2, \cdots, k$ and $j= 1,2, \cdots, k$, where $z_i$ is the $i$-th element of $\mathbf{z}$, $E_{ij}$ is the $(i,j)$-th element of matrix $\mathbf{E}$, $d_{min}$ is the smallest distance between any two points in the constellation, and $\Re$ and $\Im$ denote the real and imaginary parts respectively.
After combining the $K$ equations in \eqref{eq1} and \eqref{eq2} and taking expectations on all sides, 
we have
\begin{equation*}\label{real_ineq}
\frac{-d_{\min}}{2}\mathbf{1}_K <\mathbb{E}[\Re(\mathbf{z} - \mathbf{E}\mathbf{g})] < \frac{d_{\min}}{2}\mathbf{1}_K,
\end{equation*}
\begin{equation*}
\frac{-d_{\min}}{2}\mathbf{1}_K <\mathbb{E}[\Im(\mathbf{z} - \mathbf{E}\mathbf{g})] < \frac{d_{\min}}{2}\mathbf{1}_K.
\end{equation*}
where $\mathbf{1}_K$ is a $K\times 1$ vector with all entries as ones.

Here $\mathbb{E}(z_i) =0$, for all $i=1,2,\cdots,N$, because for a given transmitted vector $\mathbf{x}$, $\mathbf{x}_{\text{ZF}}$ can take any point in the constellation around $\mathbf{x}$ due to randomly and independently distributed noise and hence the expectation of difference between the two quantities would be zero. Therefore
\begin{eqnarray*}
\frac{-d_{\min}}{2}\mathbf{1}_K <\mathbb{E}[\Re(\mathbf{Eg})] < \frac{d_{\min}}{2}\mathbf{1}_K
\end{eqnarray*}
which, after substituting for $\mathbf{E}$ and $\mathbf{g}$, yields
\begin{eqnarray}
\frac{-d_{\min}}{2}\mathbf{1}_K <\mathbb{E}[\Re((\mathbf{\widetilde C}_k - \mathbf{C}^{-1})(\mathbf{H}^H\mathbf{H}\mathbf{x}+\mathbf{H}^H\mathbf{n}))] < \frac{d_{\min}}{2}\mathbf{1}_K \nonumber 
\end{eqnarray}
\begin{eqnarray}
\Rightarrow \frac{-d_{\min}}{2}\mathbf{1}_K <\mathbb{E}[\Re(\mathbf{S}\mathbf{x})] < \frac{d_{\min}}{2}\mathbf{1}_K, \label{final_bound_re}
\end{eqnarray}
where we define $\mathbf{S} = \mathbf{I} - \mathbf{\widetilde C}\mathbf{C}$ as the residual matrix. 
Similarly, 
\begin{equation}\label{final_bound_im}
\frac{-d_{\min}}{2}\mathbf{1}_K <\mathbb{E}[\Im(\mathbf{S}\mathbf{x})] < \frac{d_{\min}}{2}\mathbf{1}_K
\end{equation}
Hence, if \eqref{final_bound_re} and \eqref{final_bound_im} are together satisfied by $\mathbf{S}$ for a given transmitted vector $\mathbf{x}$, the ZF solutions through approximate and exact inverses would be equal. 
Next, we discuss some low complexity approximate matrix inversion methods which can be used to find ZF solution accurately. 

\subsection{Low Complexity Iterative Methods for Computing Approximate Matrix Inverses}

Several low complexity iterative methods for finding the inverse of a matrix have been proposed in \cite{iterative_method1,ref11}. 
Let $\mathbf{C}_{k}$ be the approximate inverse and $\mathbf{S}_k$ be the residual matrix after $k$ iterations. The order of the iterative method is $p$ if the residuals after $k$ and $k+1$ iterations satisfy $\mathbf{S}_{k+1} = \mathbf{S}_k^p.$ For e.g., in a third order method, approximate matrix is calculated in the following manner \cite{iterative_method1}
\begin{equation}\label{3order}
\mathbf{C}_{k+1} = \mathbf{C}_k(3\mathbf{I} - \mathbf{CC}_k(3\mathbf{I} - \mathbf{CC}_k)),
\end{equation}
where $\mathbf{I}$ is the identity matrix. Here, we note that $\mathbf{S}_{k+1} = \mathbf{S}_k^3$. Similarly, a seventh order iterative method is defined as
\begin{eqnarray}
\mathbf{C}_{k+1} &=& \mathbf{C}_k (7\mathbf{I} + \mathbf{CC}_k (−21\mathbf{I} + \mathbf{CC}_k (35\mathbf{I} + \mathbf{CC}_k (−35\mathbf{I} \nonumber\\
&+& \mathbf{CC}_k (21\mathbf{I}+ \mathbf{CC}_k (−7\mathbf{I} +\mathbf{CC}_k)))))),
\end{eqnarray}
and here, we have $\mathbf{S}_{k+1} = \mathbf{S}_k^7.$

In our simulations, we use Newton's iterative method for finding approximate matrix inverse which has low latency, low complexity \cite{ref4} and is also easy to implement \cite{ref11}. The approximate inverse is updated in each iteration according to
\begin{eqnarray}\label{newton}
\mathbf{C}_{k+1} = (2\mathbf{I} -– \mathbf{C}_k\mathbf{C})\mathbf{C}_k.
\end{eqnarray}
Here, $\mathbf{S}_{k+1} = \mathbf{S}^2_k$, revealing quadratic convergence. Increasing the number of iterations increases accuracy, but also increases the number of operations required and hence affects complexity, resulting in a trade-off between performance and efficiency.   

Initial matrix $\mathbf{C}_0$ needs to be chosen with care as it decides the number of iterations required for the method to converge, if it converges at all. The applicability of iterative methods is restricted since global convergence is not inherent to all initial matrices. A general condition for initialization is given by
$||\mathbf{I} - \mathbf{CC}_0||_2 < 1~\textrm{or}~||\mathbf{S}_0||_2 < 1$.
This condition ensures that the residual converges towards zero after each iteration.

However, there are some conventional initialization methods which guarantee convergence. In \cite{ref11}, theorem 2 shows that to find the inverse of a matrix $\mathbf{C}$, the initialization $\mathbf{C}_0 = a\mathbf{C}^H$, where $a$ satisfies $0 < a < \frac{2}{\sigma_{max}^2}$ and $\sigma_{max}^2$ is denoted as the largest eigenvalue of the matrix $\mathbf{A} = \mathbf{C}^H\mathbf{C}$, ensures convergence. To reduce the complexity, following bound is used \cite{ref11}
\begin{eqnarray}
\sigma_{max}^2 \leq \lambda_{upper} = m + t(N-1)^\frac{1}{2}
\end{eqnarray}
where $m = \frac{trace(\mathbf{A})}{N}$ and $t^2 = \frac{trace(\mathbf{A}^2)}{N}~–-~m^2$ and $a$ is selected as
$a = 2/\lambda_{upper},$    
which ensures convergence. 
In the next section, we propose a low complexity SD algorithm for large-antenna and massive MIMO systems that uses above matrix inversion methods to accurately estimate the transmitted signal vector. 

\section{Sphere Decoding using Iterative Matrix Inverse}
Now, let us investigate the advantages of using iterative matrix inverses for non-linear detectors, such as SD. Presently, there are two main versions of SD. The first is the Schnorr-Euchner enumeration \cite{Tellambura13,sesd} that updates the radius for SD adaptively, where after starting with an infinite radius, the search space shrinks with each good point until we get the optimal solution. In large/massive MIMO systems, such a technique would result in a huge decoding complexity. The other one is Fincke-Pohst algorithm based SD \cite{Hassibi05, fpsd}, which uses a fixed radius approach, and all the points that are inside the search space defined by the radius are compared for detecting the transmitted signal. This technique is extremely sensitive to the choice of the radius. It has been shown in the literature that both these approaches provide near ML performance. In this section, we propose a mechanism to reduce the complexity of SD.

Our SD algorithm combines both the strategies wherein we initialize with a BR computed using a low complexity iterative matrix inverse and also update the radius adaptively with every good point. The number of updates when using this algorithm would be significantly less, as the radius will be updated only when a new
point is closer to the transmitted signal than ZF. Also, we are always guaranteed a solution as the ZF solution is always inside the searched domain. In Algorithm \ref{DFTS-Alg}, we show the steps of the proposed SD scheme.

\begin{algorithm}[t]
\SetKwInOut{Input}{Input}\SetKwInOut{Output}{Output}\SetKwInOut{Function}{Function}
\Input{$\mathbf{y}$, $\mathbf{H}$, $\Omega$, $k$}
\Output{$\mathbf{\hat{x}}$}
\BlankLine
\emph{Initialization} $i=K$, $cost=r_k$, $\widetilde{c}_i=0$\;
$[\mathbf{Q}\,\,\,\mathbf{R}]\leftarrow \text{QR decomposition of } \mathbf{H}$ and $\mathbf{z}={\mathbf{Q}}^H\mathbf{y}$\;
$\mathbf{\hat{x}} \leftarrow \mathbf{DFTS}(\mathbf{z},\,\mathbf{R},\,\Omega,\,cost,\,\widetilde{c}_i,\,d,\,i)$\;
\BlankLine
\Function{$\mathbf{DFTS}(\mathbf{z},\,\mathbf{R},\,\Omega,\,cost,\,\widetilde{c}_i,\,i)$}

\For{$j\leftarrow 1$ \KwTo length$(\Omega)$}{
${c}_j=|z_{i}-r_{i,i}x_j|^2$, $\forall x_j\in\Omega$\;
}
Sort $c_j$'s in ascending order and keep only those symbols for which $c_j< (cost-\widetilde{c}_i)$\;

\eIf{$c_j\nleq (cost-\widetilde{c}_i)$}{
\Return $\mathbf{\hat{x}}$, $cost$\;
}{
\For{$u\leftarrow 1$ \KwTo length$(\mathbf{c})$}{
$\hat{x}_i=x_u$\;
$\widetilde{c}_i\leftarrow \widetilde{c}_i+c_u$\;
\eIf{$i=1$}{
\If{$cost_{temp}<cost$}{
$cost\leftarrow \widetilde{c}_i$\;
\Return $\mathbf{\hat{x}}$, $cost$\;
}
}{
$\mathbf{\widetilde{z}}=\mathbf{z}-\mathbf{R}_{:,u}\,x_u$\;
Extend the tree $\mathcal{T}$ for all $\Omega$\;
$[\mathbf{\hat{x}},cost]\leftarrow \mathbf{DFTS}(\mathbf{\widetilde{z}},\mathbf{R},\Omega,cost,\widetilde{c}_i,i-1)$\;
}
}
}
\caption{Proposed SD Scheme}
\label{DFTS-Alg}
\end{algorithm}

\subsection{Comparison of Babai Radii Calculated through Approximate and Exact Matrix Inverses}

Though iterative methods provide a good approximate inverse, it is important to analyze the effect of approximation on the
BR, as the choice of radius largely governs the complexity of
SD. Interestingly, we show that the application of approximate
matrix inversion methods also reduces the value of radius
which leads to further savings in complexity.

To prove this, let us define $r_e$ as the BR computed through
exact inverse and $r_k$ as the BR computed through the iterative method after $k$ iterations.
We know that $r_e = \lim_{k\rightarrow\infty} r_k$. Now,
from the definition of BR \cite{Hassibi05, babai2}, we can write

\begin{eqnarray}\label{r_e}
r_e &=& ||\mathbf{R}(\mathbf{x}_{\text{ZF}} - \mathbf{\hat{x}}_1)||,\label{r_e}\\
r_k &=& ||\mathbf{R}(\mathbf{x}_{\text{ZF}} - \mathbf{\hat{x}}_2)||,\label{r_k}
\end{eqnarray}
where $\mathbf{\hat{x}}_1 = \mathbf{C}^{-1}\mathbf{g}$, $\mathbf{\hat{x}}_2 = \mathbf{C}_k\mathbf{g}$ and $\mathbf{R}$ is obtained from the QR decomposition of $\mathbf{H}$ as $\mathbf{H} = \mathbf{QR}$. Let $\mathbf{\bar{n}}$ denote the noise with respect to $\mathbf{x}_{\text{ZF}}$, i.e.,
\begin{eqnarray}\label{receiver_sys_model}
\mathbf{y}= \mathbf{H}\mathbf{x}_{\text{ZF}} + \mathbf{\bar{n}}.
\end{eqnarray} 
Now, let us define the relation between transmitted vector $\mathbf{x}$ and detected ZF vector $\mathbf{x}_{\text{ZF}}$ as
\begin{equation}\label{delta}
\mathbf{x} =\mathbf{x}_{\text{ZF}} + \Delta,
\end{equation} 
where $||\Delta||$ denotes the magnitude of error in $\mathbf{x}_{\text{ZF}}$. 
Since $\mathbf{x}$ and $\mathbf{x}_{\text{ZF}}$ both belong to the same constellation, expectation of the difference between $\mathbf{x}$ and $\mathbf{x}_{\text{ZF}}$ would be zero. Substituting \eqref{delta} in \eqref{receiver_sys_model}, we get
$\mathbf{\bar{n}} = \mathbf{n} - \mathbf{H}\Delta,$
and thus $\mathbb{E}(\mathbf{\bar{n}}) = 0$.

\begin{figure}
\centering
\includegraphics[width=\columnwidth, height=0.65\columnwidth]{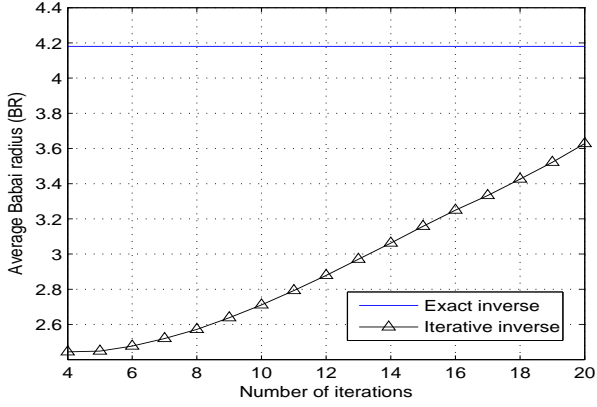}
\caption{BR with Newton's iterative method for a $16 \times 16$ system.}
\label{fig1}
\end{figure}

For a sufficient number of iterations, we can write the expected difference between the squares of the two radii in (\ref{r_e}) and (\ref{r_k}) as
\begin{eqnarray}\label{afterapprox}
\mathbb{E}[r_e^2 - r_k^2] = 2\Re[\mathbb{E}\{\mathbf{\bar{n}}^H\mathbf{H}\mathbf{S}_k\mathbf{C}^{-1}\mathbf{g}\}].
\end{eqnarray}
We prove the above equation in Appendix I.

\begin{figure}[t]
\centering
\includegraphics[width=\columnwidth, height=0.65\columnwidth]{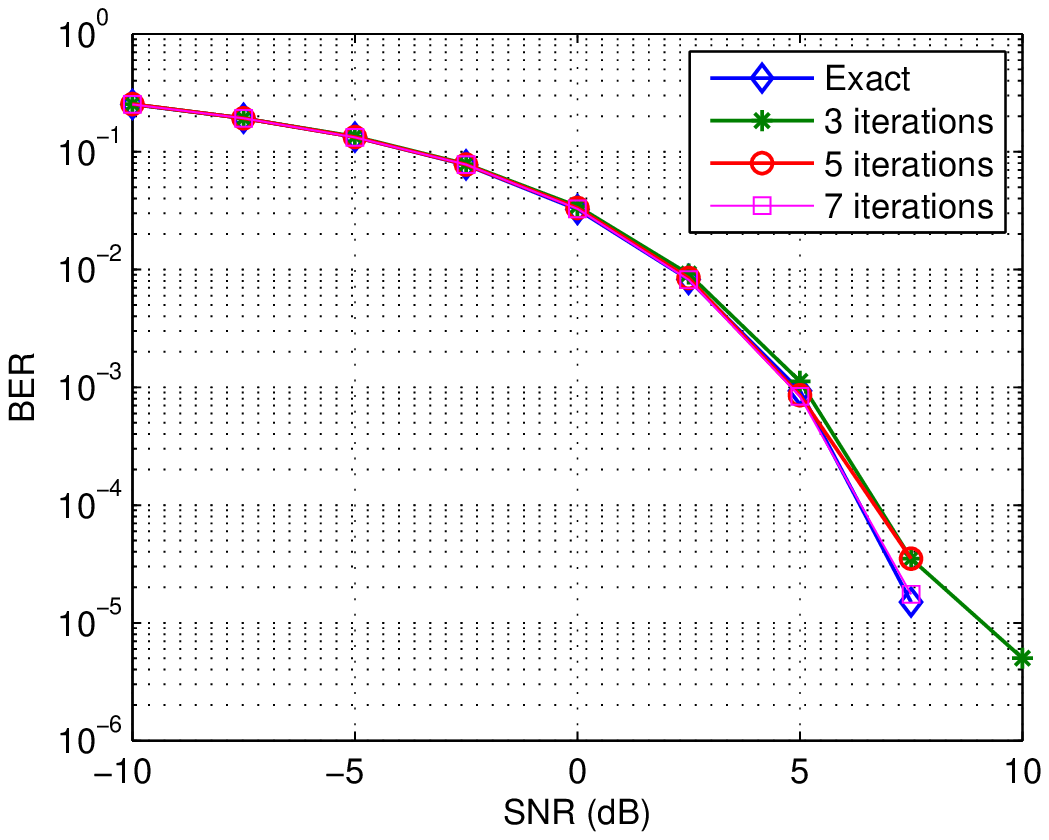}
\caption{Bit error performance for the MMSE decoder in a massive MIMO system with $N=128$, $K=8$ for 16-QAM.}
\label{fig2}
\end{figure}

We next show that the L.H.S. in \eqref{afterapprox} decreases as the number of iterations increase.
Using $\mathbf{g} = \mathbf{H}^H\mathbf{y}$ in \eqref{afterapprox}, we can write
\begin{eqnarray}\label{Eeverywhere}
\nonumber \mathbb{E}[\mathbf{\bar{n}}^H\mathbf{H}\mathbf{S}_k\mathbf{C}^{-1}\mathbf{g}] = \mathbb{E}[\mathbf{\bar{n}}^H\mathbf{H}\mathbf{S}_k\mathbf{C}^{-1}\mathbf{H}^H(\mathbf{H}\mathbf{x}_{\text{ZF}} + \mathbf{\bar{n}})] \\
 = \mathbb{E}[\mathbf{\bar{n}}^H\mathbf{H}\mathbf{S}_k\mathbf{x}_{\text{ZF}}] + \mathbb{E}[\mathbf{\bar n}^H\mathbf{H}\mathbf{S}_k\mathbf{C}^{-1}\mathbf{H}^H\mathbf{\bar{n}}].
\end{eqnarray}
Also, for a given channel matrix $\mathbf{H}$ and received vector $\mathbf{y}$,  $\mathbf{x}_{\text{ZF}}$ would be a constant. Therefore, we can take vector $\mathbf{x}_{\text{ZF}}$ out of the first expectation term in (\ref{Eeverywhere}) and it becomes
\begin{eqnarray}
\mathbb{E}[\mathbf{\bar{n}}^H\mathbf{H}\mathbf{S}_k\mathbf{x}_{\text{ZF}}] = \mathbb{E}[\mathbf{\bar{n}}^H]\mathbf{H}\mathbf{S}_k\mathbf{x}_{\text{ZF}} = 0,
\end{eqnarray}
and therefore, can be rewritten as
\begin{eqnarray}\label{tracesk}
\mathbb{E}[\mathbf{\bar n}^H\mathbf{H}\mathbf{S}_k\mathbf{C}^{-1}\mathbf{g}] &=& \mathbb{E}[\mathbf{\bar n}^H\mathbf{H}\mathbf{S}_k\mathbf{C}^{-1}\mathbf{H}^H\mathbf{\bar{n}}]\nonumber\\&=& N_0\text{Tr}(\mathbf{H}\mathbf{S}_k\mathbf{C}^{-1}\mathbf{H}^H),
\end{eqnarray}
where Tr$(\mathbf{X})$ denotes the trace of matrix $\mathbf{X}$. From \eqref{afterapprox} and \eqref{tracesk}
\begin{eqnarray}\label{Ek}
\mathbb{E}[r_e^2 - r_k^2] &=& 2N_0\Re[\text{Tr}(\mathbf{H}\mathbf{S}_k\mathbf{C}^{-1}\mathbf{H}^H)]\nonumber\\
&=& 2N_0\Re[\text{Tr}(\mathbf{S}_k\mathbf{C}^{-1}\mathbf{H}^H\mathbf{H})] \nonumber \\
&=& 2N_0\Re[\text{Tr}(\mathbf{S}_k)].
\end{eqnarray}
Similarly, for the radius obtained after $k+1$ iterations, we get
\begin{eqnarray}\label{Ek+1}
\mathbb{E}[r_e^2 - r_{k+1}^2] = 2N_0\Re[\text{Tr}(\mathbf{S}_{k+1})].
\end{eqnarray}
It can be seen that for the residual matrix $\mathbf{S}_k = \mathbf{I} - \mathbf{C}_k\mathbf{C}$, we have $\text{Tr}(\mathbf{S}_k) \geq 0$. If the iterative methods used for matrix inversion converges to the exact inverse, it can be assumed that 
$\text{Tr}(\mathbf{S}_k) > \text{Tr}(\mathbf{S}_{k+1})$,
as the elements of the residual matrix will tend towards zero as the number of iterations increase. Therefore, from equations (\ref{Ek}) and (\ref{Ek+1}), it can be deduced that \begin{eqnarray*}
\mathbb{E}[r_e^2 - r_{k+1}^2] &<& \mathbb{E}[r_e^2 - r_k^2]\\
\Rightarrow \mathbb{E}[r_{k+1}^2] &>& \mathbb{E}[r_k^2]
\end{eqnarray*}
which means that, in general, BR after $k$ iterations is smaller than the BR calculated after $k+1$ iterations. In Fig. \ref{fig1}, we use Newton's iterative method for computing the approximate inverse and plot the BR for different iterations for a $16 \times 16$ MIMO system. A monotonic rise in the value of BR with increasing iterations corroborates the above analysis.

As $r_e = \lim_{k\rightarrow\infty} r_k$, therefore 
$$\mathbb{E}[r_{e}^2] > \mathbb{E}[r_k^2] ~~\text{for finite $k$},$$
i.e. the BR calculated using the exact inverse is larger than the BR calculated through an iterative method for all the iterations.
Thus, as stated before, an approximate inverse can provide twofold savings in complexity.

\begin{figure}
\centering
\includegraphics[width=\columnwidth, height=0.65\columnwidth]{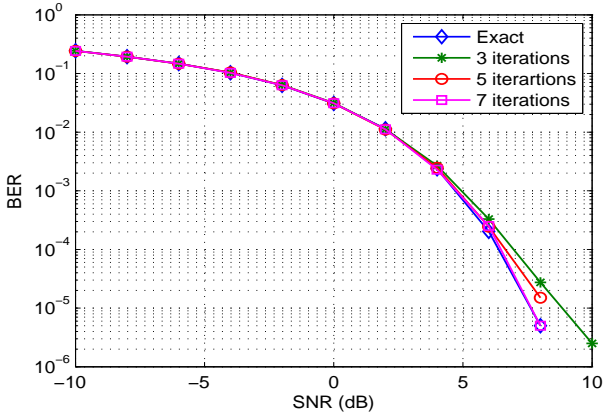}
\caption{Bit error performance for the ZF decoder in a massive MIMO system with $N=128$, $K=8$ for 16-QAM.}
\label{fig3}
\end{figure}

\begin{figure}
\centering
\includegraphics[width=\columnwidth, height=0.65\columnwidth]{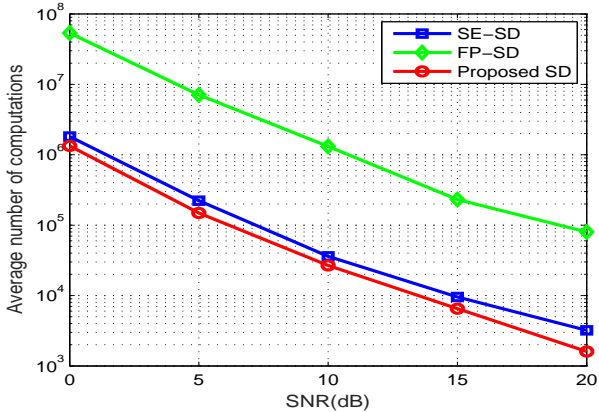}
\caption{Average number of computations for different SD schemes for a $16 \times 16$ large MIMO system for 4-QAM.}\label{fig_16x16}
\end{figure}


\section{Simulation Results}


We first examine the performance of ZF and MMSE detectors for massive MIMO scenarios. Subsequently, we compare the simulation results for different SD methods that exist in the literature to the scheme proposed in Algortihm \ref{DFTS-Alg}. In Fig. \ref{fig2} and Fig. \ref{fig3}, we plot Bit-Error-Rates (BER) for MMSE and ZF decoders, respectively, for the cases when the matrix inverse is calculated exactly and using Newton's iterative method. We calculate the approximate inverse for 3, 5 and 7 iterations. We see that for 3 and 5 iterations, the error performance in the case of MMSE is slightly away from the case when the exact inverse is used. However, increasing the number of iterations to 7 provides identical performance. Using more number of iterations would not result in any performance gain. Similarly, in the case of ZF decoding, performance improves with the number of iterations, and 7 iterations provides the same performance as the ZF decoder using the exact inverse. 

\begin{figure}[t]
\centering
\includegraphics[width=\columnwidth, height=0.65\columnwidth]{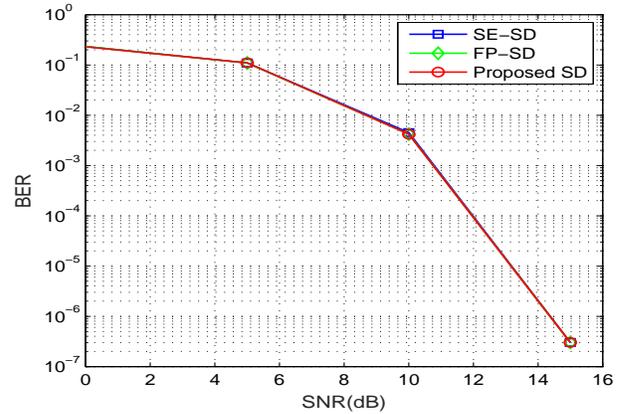}
\caption{Bit error performance for different SD schemes for a $16 \times 16$ large MIMO system for 4-QAM.}
\label{fig_16x16ber}
\end{figure}

\begin{figure}
\centering
\includegraphics[width=\columnwidth, height=0.65\columnwidth]{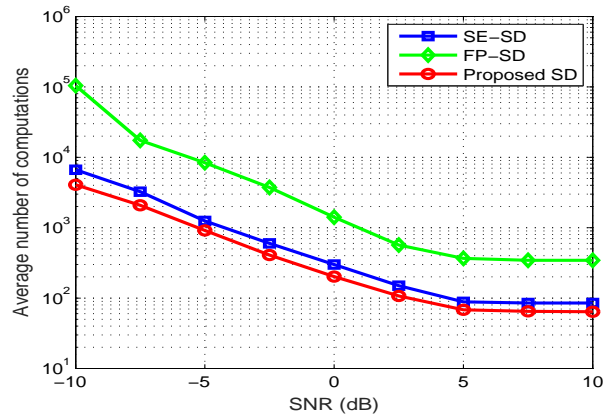}
\caption{Average number of computations for different SD schemes for a massive MIMO system with 32 base antennas and 8 users for 4-QAM.}
\label{fig_32x8}
\end{figure} 

\begin{figure}
\centering
\includegraphics[width=\columnwidth, height=0.65\columnwidth]{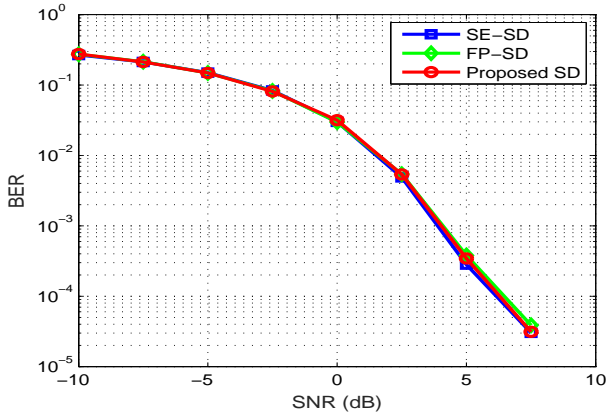}
\caption{Bit error performance for different SD schemes for a massive MIMO system with 32 base antennas and 8 users for 4-QAM.}
\label{fig_32x8ber}
\end{figure}

We also perform Monte Carlo simulations for BER and average number of computations for the three different SD schemes discussed above. The first two are adaptive radius (SE-SD) and fixed radius (FP-SD) algorithms respectively. We compare these conventional schemes with the SD scheme proposed in Algorithm \ref{DFTS-Alg} in terms of performance and average number of computations required to find the solution. We use Newton's iterative method with 7 iterations to calculate the approximate matrix inverse. In Fig. \ref{fig_16x16}, we compare the average number of computations required by the three schemes for a $16 \times 16$ system. It can be observed from the figure that the proposed SD scheme takes at least 35\% less number of computations compared to the other two schemes. Also, from Fig. \ref{fig_16x16ber}, we can deduce that there is no reduction in the quality of performance as all the three schemes give the same BER. In Fig. \ref{fig_32x8} and Fig. \ref{fig_32x8ber}, we present similar numerical results for an $N = 32$ and $K= 8$ massive MIMO system. We again note that our SD scheme outperforms the conventional scheme while providing the same error performance. 

\section{Conclusion}
We have shown the advantages of using an approximate matrix inverse for detectors in large/massive MIMO systems. We obtained the maximum error which can be tolerated in the inverse to arrive at the same quantized ZF/MMSE solution. Simulation results show that iterative inversion methods, used to calculate the ZF and MMSE solutions, reached the same performance as provided by the exact inverse for sufficient number of iterations. Extending the idea to complex detectors like SD, we show that the value of BR calculated using iterative methods is less than the BR obtained through the exact method. To this end, we proposed an adaptive SD scheme that uses BR as the initial radius. Simulation results show that the proposed SD scheme outperforms FP-SD and SE-SD in terms of complexity without any loss in performance.

\section*{Appendix}

To prove \eqref{afterapprox}, we use the definition of $r_e$ and $r_k$ from \eqref{r_e} and \eqref{r_k} so that
\begin{eqnarray*}\label{main_eq}
r_e^2 - r_k^2 &=& ||\mathbf{R}(\mathbf{\hat{x}}_1-\mathbf{x}_{\text{ZF}})||^2 - ||\mathbf{R}(\mathbf{\hat{x}}_2-\mathbf{x}_{\text{ZF}})||^2\\
 &=& ||\mathbf{RC}^{-1}\mathbf{g}||^2 - ||\mathbf{RC}_k\mathbf{g}||^2 \\
&+& 2\{\Re[(\mathbf{Rx}_{\text{ZF}})^H\mathbf{R}(\mathbf{C}_k - \mathbf{C}^{-1})\mathbf{g}]\}
\end{eqnarray*}
Now, using the fact that $\mathbf{C}_k = \mathbf{C}^{-1}+\mathbf{E}_k$, 
 we get
\begin{eqnarray}
r_e^2 - r_k^2 = 2\Re[(\mathbf{x}_{\text{ZF}} - \mathbf{C}^{-1}\mathbf{g})^H\mathbf{R}^H\mathbf{RE}_k\mathbf{g}] - ||\mathbf{RE}_k\mathbf{g}||^2.
\end{eqnarray}
After using \eqref{receiver_sys_model} and taking expectations on both sides, we get
{\small
\begin{eqnarray}\label{re2-rk2}
\mathbb{E}[r_e^2 - r_k^2] = \mathbb{E}[2\Re\{(-\mathbf{C}^{-1}\mathbf{H}^H\mathbf{\bar n})^H\mathbf{R}^H\mathbf{RE}_k\mathbf{g}\} - ||\mathbf{RE}_k\mathbf{g}||^2].
\end{eqnarray}}
We will be neglecting the second term in R.H.S. of (\ref{re2-rk2}) citing the following assertion  
\begin{eqnarray}\label{reky}
||\mathbf{RE}_k\mathbf{g}||^2 = (\mathbf{RE}_k\mathbf{g})^H(\mathbf{RE}_k\mathbf{g}) = \mathbf{g}^H\mathbf{E}_k^H\mathbf{R}^H\mathbf{R}\mathbf{E}_k\mathbf{g}. 
\end{eqnarray}
From the orthogonal property of $\mathbf{Q}$, we have $\mathbf{R}^H\mathbf{R} = \mathbf{H}^H\mathbf{H} = \mathbf{C}$
and therefore (\ref{reky}) becomes
\begin{eqnarray}
||\mathbf{RE}_k\mathbf{g}||^2 &=& \mathbf{g}^H\mathbf{E}_k^H\mathbf{C}\mathbf{E}_k\mathbf{g} \nonumber 
\end{eqnarray}
Using $\mathbf{E}_k = \mathbf{C}_k - \mathbf{C}^{-1}$, we have
\begin{eqnarray*}
||\mathbf{RE}_k\mathbf{g}||^2 &=& \mathbf{g}^H(\mathbf{C}_k^H - (\mathbf{C}^{-1})^H)\mathbf{C}(\mathbf{C}_k - \mathbf{C}^{-1})\mathbf{g}\\
&=& \mathbf{g}^H(\mathbf{C}_k^H\mathbf{C}^H - \mathbf{I})(\mathbf{C}_k\mathbf{C} - \mathbf{I})\mathbf{C}^{-1}\mathbf{g}\\
&=& \mathbf{g}^H\mathbf{S}_k^H\mathbf{S}_k\mathbf{\hat x_1}, 
\end{eqnarray*}
where $\mathbf{S}_k = \mathbf{I} - \mathbf{C}_k\mathbf{C}$ is the residual matrix. Here, we have used the fact that $\mathbf{C}$ is a Hermitian matrix and $(\mathbf{C}^{-1})^H = \mathbf{C}^{-1}$.
Also, the first term in the R.H.S. of (\ref{re2-rk2}) can be written as
\begin{eqnarray*}
\mathbb{E}[2\Re\{(-\mathbf{C}^{-1}\mathbf{H}^H\mathbf{\bar{n}})^H\mathbf{R}^H\mathbf{RE}_k\mathbf{g}\}] 
= \mathbb{E}[2\Re(\mathbf{\bar{n}}^H\mathbf{HE}_k\mathbf{g})]\\
= \mathbb{E}[2\Re(\mathbf{\bar{n}}^H\mathbf{H}\mathbf{S}_k\mathbf{C}^{-1}\mathbf{g})].
\end{eqnarray*}
For sufficient number of iterations, $\mathbf{S}_k$ would be very small and hence the term $\mathbb{E}[||\mathbf{RE}_k\mathbf{g}^2||]$ can be neglected when compared to first term in the R.H.S. of equation (\ref{re2-rk2}), as the former is proportional to $\mathbf{S}_k^H\mathbf{S}_k$ while the latter is proportional to $\mathbf{S}_k$. 
Hence, (\ref{re2-rk2}) can be rewritten as
\begin{eqnarray*}
\mathbb{E}[r_e^2 - r_k^2] &\approx& \mathbb{E}[2\Re(\mathbf{\bar{n}}^H\mathbf{H}\mathbf{S}_k\mathbf{C}^{-1}\mathbf{g})]
\end{eqnarray*}
which proves \eqref{afterapprox} for sufficient number of iterations $k$.

\bibliographystyle{IEEEtran}
\bibliography{IEEEabrv,References_ICCw}
\end{document}